\documentclass[preprint]{aastex}

\slugcomment{to appear in The Astronomical Journal}  

\begin{document}

\title{Search for Molecular Gas in the quasar SDSS~1044-0125 at z=5.73}

\author{D.J. Wilner}
\affil{Harvard-Smithsonian Center for Astrophysics, 60 Garden Street,
    Cambridge, MA 02138}
\email{dwilner@cfa.harvard.edu}
\and
\author{M.C.H. Wright and J. Di~Francesco}
\affil{Radio Astronomy Laboratory, University of California at Berkeley,
Campbell Hall, Berkeley, CA 94720}
\email{wright@astron.berkeley.edu,jdifran@astron.berkeley.edu}

\begin{abstract}
We report upper limits on CO J=2--1 and CO J=5--4 emission 
from the quasar SDSSp J104433.04-012502.2 at $z=5.73$ from 
observations made with the Berkeley-Illinois-Maryland-Association Array. 
Previously reported limits on CO J=6--5 emission (Iwata et al. 2001)
were obtained at $z=5.80$, which is now thought to be off by 1\%, 
and the observations likely missed the relevant redshifts for molecular gas.
The new $3\sigma$ upper limits on the line luminosities are 
$L^{'}_{CO}(2-1) < 5.1 \times 10^{10}$ K~km~s$^{-1}$~pc$^2$ 
and 
$L^{'}_{CO}(5-4) < 3.0 \times 10^{10}$ K~km~s$^{-1}$~pc$^2$,
assuming 200~km~s$^{-1}$ linewidth. 
The CO J=5--4 observations place an upper limit on 
warm, dense molecular gas mass comparable to amounts 
derived for some other high redshift quasar systems
from detections of this line.  The limit on CO J=2--1 emission 
suggests that excitation bias does not affect this conclusion.  
In addition, no molecular gas rich companion galaxies 
are found in a $\sim1.4$~Mpc field surrounding the quasar.
\end{abstract}


\section{Introduction}
The study of the star formation properties and gas content
of galaxies at far cosmological distances is one important step
toward understanding galaxy formation and evolution.
The quasar SDSSp~J104433.04-012502.2 at $z\approx5.8$ 
(hereafter SDSS~1044-0125), 
discovered by Fan et al. (2000) using the Sloan Digital Sky Survey, 
is among the highest redshift objects known. 
Recent observations of this quasar with SCUBA on the
JCMT at 850~$\mu$m detect thermal continuum emission (reported by
Iwata et al. 2001), which suggests a large reservoir of dust 
and therefore also molecular gas. 
The presence of a substantial gas mass gains support 
from the apparent X-ray weakness of the quasar, which 
likely results from heavy intrinsic absorption 
(Brandt et al. 2001, Mathur 2001).
Molecular gas has been detected from 
at least a dozen $z>2$ objects through CO lines at millimeter wavelengths, 
and these observations provide important clues to the formation history 
of galaxies and their relationship to supermassive black holes.

A recent search for CO J=6--5 emission from SDSS~1044-0125 
by Iwata et al. (2001) reported an upper limit on the inferred 
molecular gas mass comparable 
to the detections for some high redshift quasars. 
Unfortunately, the search was centered at $z=5.80$,
the initial redshift estimated by Fan et al. (2000).
Recent spectroscopic studies of SDSS~1044-0125 give a 
more accurate value of $z=5.73\pm0.01$ (Djorgovski et al. 2001,
see also Goodrich et al. 2001), about 1\% off from the initial estimate.
Because of the narrow instantaneous bandwidth available to 
current millimeter interferometers, the revised redshift falls 
outside window that was searched for CO J=6--5 emission, and
the observations likely missed the relevant redshifts for 
molecular gas in the quasar host.  
Since SDSS~1044-0125 has several properties in common with 
high redshift quasars where CO emission has been detected,
(enumerated by Iwata et al. 2001), 
the revision of the optical redshift determination gives impetus to 
a new search for molecular gas.

A potentially important limitation of searching for emission from 
CO lines with high rotational quantum numbers, like the J=6--5 transition, 
is that prevailing physical conditions may be insufficiently extreme 
to excite these lines.  The surprising detection of extended 
emission in the low excitation CO J=2--1 line towards the 
quasar APM~08279+5255 at $z=3.91$ (Papadopoulos et al. 2001)
suggests that low excitation CO lines can reveal molecular mass reservoirs 
that are one or two orders of magnitude larger than suggested by 
observations of high excitation CO lines. 

In this short paper, we present results of searches for CO J=2--1 and J=5--4 
emission from SDSS~1044-0125 using the Berkeley-Illinois-Maryland Array 
(BIMA)\footnote{The BIMA array is operated by
the Berkeley-Illinois-Maryland Association under funding from the
National Science Foundation.} (Welch et al. 1996)
that provide new limits on the amount of molecular gas 
associated with this luminous high redshift quasar.  
The BIMA 1~cm band receiver system, which was developed primarily for 
observations of the Sunyaev-Zeldovich effect (Carlstrom, Joy \& Grego 1996, 
Grego et al. 2000), provides a unique facility to search for 
highly redshifted low lying CO lines. For SDSS~1044-0125, 
the CO J=2--1 line is redshifted to 34 GHz, within the accessible 
tuning range. The standard digital correlator allows for several times 
larger velocity coverage than generally available at shorter wavelengths, 
sufficient to span the uncertainty in the quasar redshift determined from 
optical lines, as well as the typical kinematic offsets of molecular gas 
from the redshift derived optically. In addition, the small BIMA array 
antennas provide a large field of view, which enables
imaging the quasar environs over Mpc scales at 34 GHz in a single pointing.

\section{Observations}

\subsection{CO J=2--1 Line}
Observations of the CO J=2--1 line 
(redshifted frequency $230.5380/(1+5.73)= 34.255$~GHz)
were conducted in two parts, in September 2000 and September 2001. 
Both sets of observations used the 9 antennas of the BIMA array 
equipped with 1~cm band receivers. 
The pointing center was $\alpha(J2000)=10^{h}44^{m}33\fs04$,
                        $\delta(J2000)=-01^{\circ}25\farcm02\farcs2$.
The 2000 observations, made with the correlator centered at 
the nominal redshift $5.80$ estimated by Fan et al. (2000),
were obtained on 14 days, generally in short tracks of less than 
a few hours near source transit. The array antennas were in C configuration, 
giving $\sim25''$ resolution.  
The 2001 observations, made with the correlator 
centered at the revised redshift $5.73$ determined by 
Djorgovski et al. (2001), were obtained on 2 days in longer tracks.
The array antennas were in B configuration, giving $\sim8''$ resolution. 
For both sets of observations, the hybrid correlator was configured 
with 8 individual windows, each with 32 channels spanning 100 MHz,
in some cases overlapped to avoid gaps at the window edges. 
The system temperatures ranged from about 50 to 100 K (SSB).  
Short observations of the nearby calibrator J1058+015 were 
made every half hour to track the interferometer phase
and also for bandpass calibration.
The flux scale was determined through observations of Mars that were
used to set (constant) {\em a priori} gains for the individual array 
antennas; the minimal amplitude drifts of the system at this low frequency 
suggest better than 10\% accuracy.

All data calibration and imaging were performed using standard routines 
in the Miriad software package. The visibility data were imaged 
with both natural and system temperature weighting for maximum sensitivity.
Since some frequencies were observed in both sets of observations, 
and some were observed in only one or the other, there are 
three regimes for the synthesized beam sizes and rms noise levels.
The frequency coverage and parameters of the resulting images are
as follows, where the rms is quoted for a 200~km~s$^{-1}$ velocity bin
($\sim8$ channels), a resolution appropriate 
for the linewidths of galactic potentials:
(1) 33.82 to 34.23 GHz observed in both 2000 and 2001, 
beam $24''\times15''$ p.a. 3$^{\circ}$, rms 0.86 mJy,
(2) 33.48 to 33.82 GHz observed in 2000 only, 
beam $35''\times18''$ p.a. 7$^{\circ}$, rms 1.00 mJy,
and (3) 34.23 to 34.55 GHz observed in 2001 only, 
beam $11''\times8''$ p.a. -15$^{\circ}$, rms 1.67 mJy.
Note that the best sensitivity was achieved for the velocity range 
redshifted by up to a few thousand km~s$^{-1}$ from the (accurate) 
optical redshift-- the range most likely to contain CO emission.
The typical kinematic offset of molecular emission in high redshift
quasar hosts from optical lines is $\sim500$~km~s$^{-1}$, with the 
largest known offset about $2500$~km~s$^{-1}$ (in APM~08279+5255, 
Downes et al. 1999). 

\subsection{CO J=5--4 Line}
Observations of the CO J=5--4 line 
(redshifted frequency $576.2679220/(1+5.73) = 85.627$~GHz)
were conducted on two days in October 2001. These observations 
used 10 antennas in D configuration, giving $\sim20''$ resolution.  
As for the September 2001 observations with the 1~cm receiver system, 
the correlator was configured to span nearly 800 MHz centered near 
the revised redshift of $z=5.73$.
The system temperatures ranged from 150 to 600 K (SSB).  
Frequent observations of the nearby calibrator J1058+015 were used
to track amplitude and phase, and bandpass calibration was checked 
with short observations of the strong source 3C273.
The flux density scale was derived from observations of Mars
and should be accurate to 20\%.
The frequency coverage and parameters of the resulting images are:
85.22 to 85.92 GHz, beam $26''\times20''$ p.a. 20$^{\circ}$, 
rms 3.0 mJy, for 200~km~s$^{-1}$ resolution.

\section{Results and Discussion}

\begin{figure}
\epsscale{0.9}
\plotone{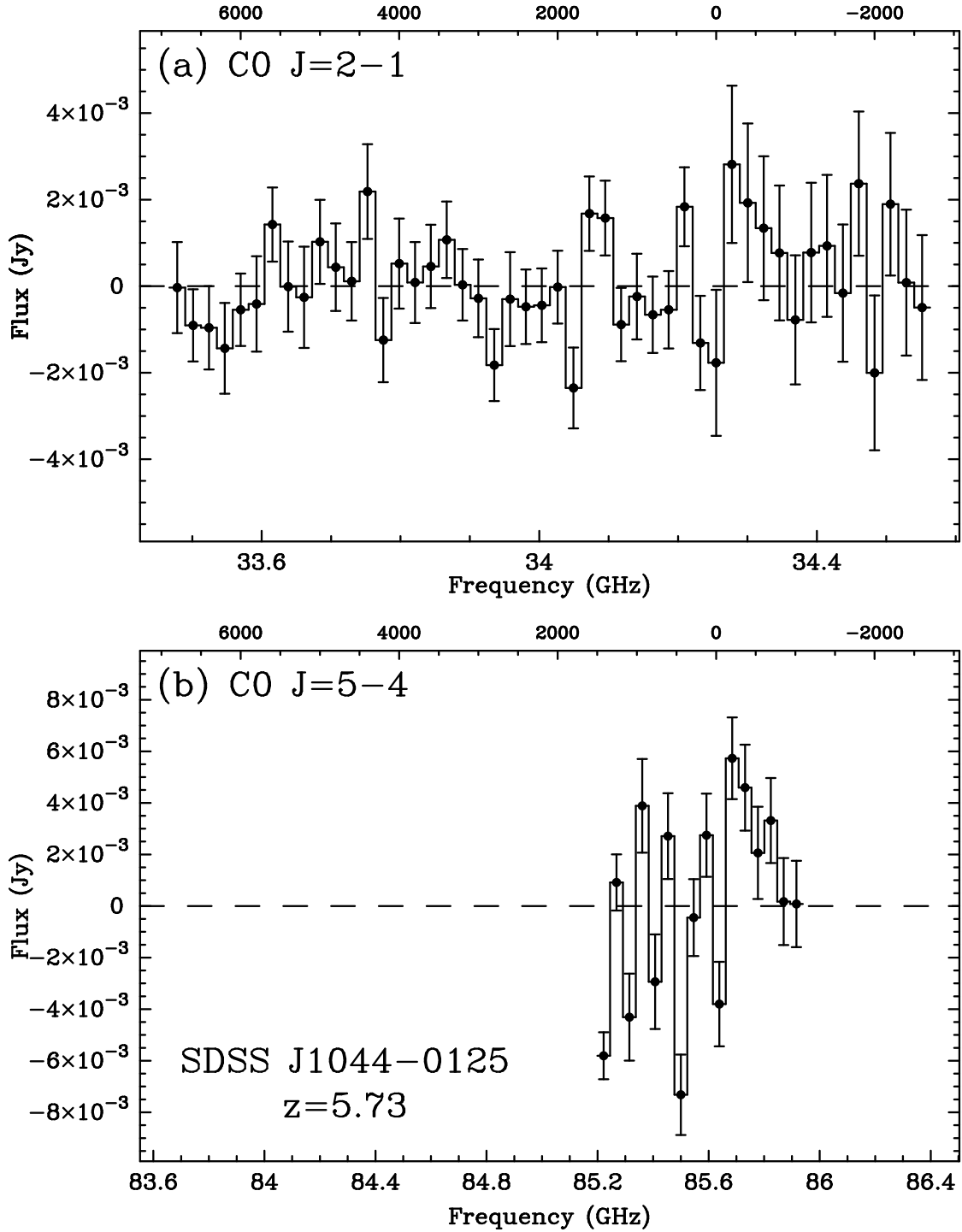}
\caption{
Spectra of (a) CO J=2--1 and (b) CO J=5--4 line emission at the 
position of SDSS~1044-0125. No significant emission is detected. 
Errorbars shown are $\pm1\sigma$ derived from the images.
The scale at the top of each panel indicate velocity (in km~s$^{-1}$) 
with respect to the nominal redshift of $5.73$. 
}
\label{fig:spectrum}
\end{figure}

\begin{figure}
\epsscale{1.0}
\plotone{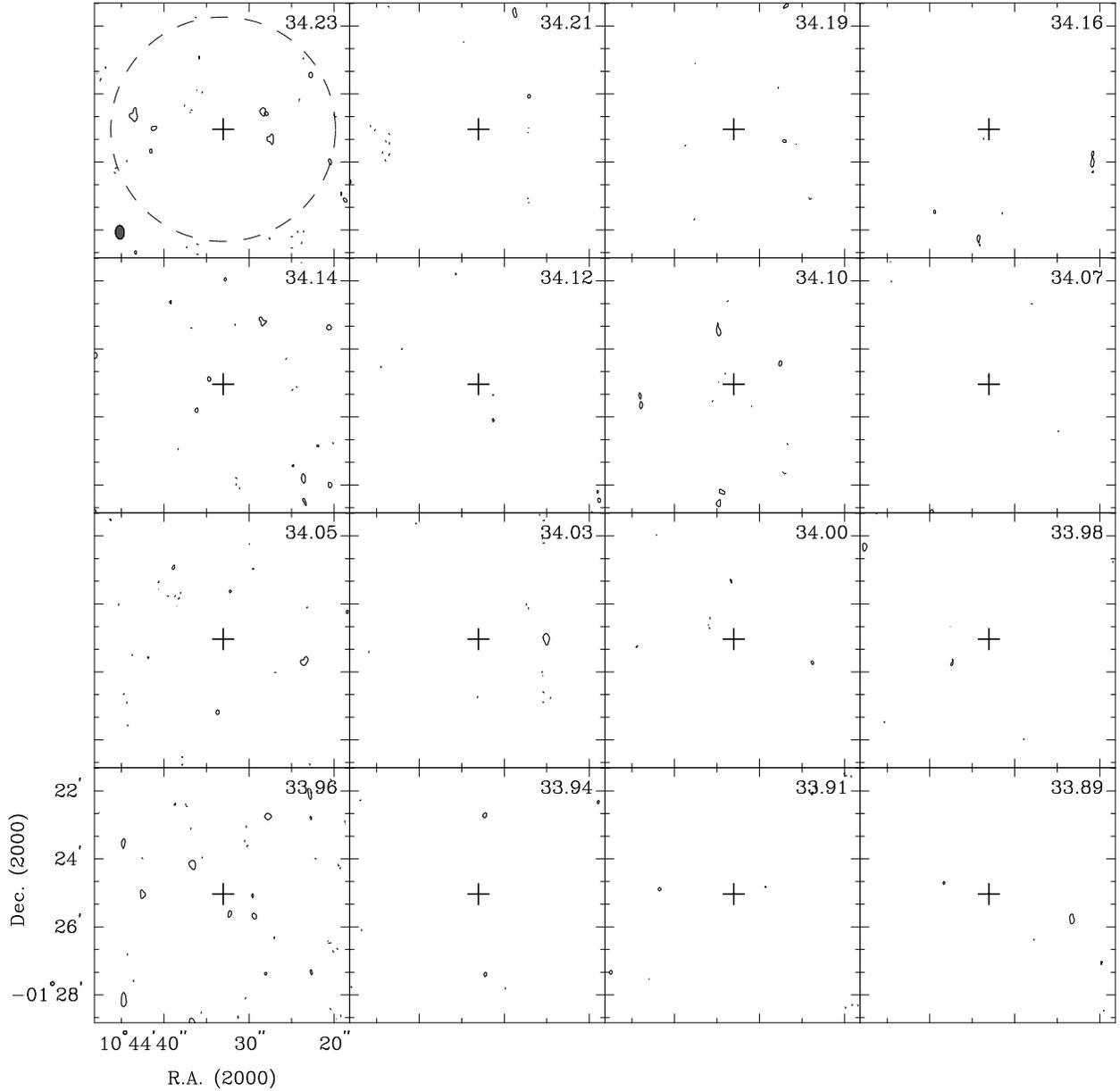}
\caption{
Channel maps for the CO J=2--1 line observations over 
the full $6\farcm$6 primary beam half power field of view 
(dashed circle). The cross in each panel marks the quasar position. 
The frequency for each panel is noted in its upper right corner.
The contour levels are $\pm3,5\times0.86$~mJy.
Negative contours are dotted.
The upper left panel shows the synthesized beam of 
$24''\times15''$ in its lower left corner.
}
\label{fig:channels}
\end{figure}

Figure~\ref{fig:spectrum} shows the CO J=2--1 and CO J=5--4 spectra 
obtained at quasar position. 
The velocity binnings for the two lines are 200 and 162~km~s$^{-1}$, 
respectively. 
The noise in the CO J=2--1 spectrum is not uniform, and empirical 
$\pm1\sigma$ error bars derived from the rms noise measured from the images 
are shown for each velocity bin.  There are some tantalizing hints of signal 
in adjacent channels close to the expected velocities, but features with 
similar (low) significance are present elsewhere in the data, and 
we do not consider any of these features to be reliable line detections.
Figure~\ref{fig:channels} shows a series of maps with 200~km~s$^{-1}$ width 
that span the full half power field of view ($6\farcm6$) for the 
J=2--1 line. 
Various attempts at smoothing in both space and frequency did not 
uncover any significant CO emission in either of the observed transitions.

Two mechanisms have been suggested for heating large masses of dust 
and gas in high redshift quasar systems: (1) high energy photons
emitted from gases accreted onto a massive black hole and 
(2) bursts of star formation. If the dust is heated by the activity
of a massive black hole, 
then bright emission may be expected from high excitation CO lines in a 
compact region close to the exciting source, from large amounts of warm,
dense gas involved in fueling and accretion. 
The CO J=5--4 line, whose upper energy level lies 
88~K above the ground state, 
requires warm gas ($>30~K$) at high densities ($>10^3$~cm$^{-3}$) 
to be populated significantly by H$_2$ collisions. Consequently, 
the upper limits on CO emission from the J=5--4 line
constrains primarily the amount of molecular gas 
with these conditions close to the massive black hole or other powerful 
heating sources. On the other hand, such extreme physical conditions 
are not necessarily appropriate for starbursts, which are likely to be 
distributed over larger spatial scales and involve cooler, 
more diffuse molecular gas. 
If the dust is heated by primarily by star formation, then emission in
CO J=2--1 line may be a more appropriate tracer of molecular gas content, 
given that the upper energy level lies just 16~K above ground and the 
excitation requirements are significantly less stringent. 

Following Solomon, Downes \& Radford (1992), we calculate upper limits 
to CO line luminosities with the expression
\begin{equation}
L^{'}_{CO} = 3.25\times10^{7} 
             S_{CO} \Delta v \nu_{obs}^{-2} D_L^2 (1+z)^{-3}~~~~
             {\rm K~km~s}^{-1}~{\rm pc}^2, 
\end{equation}
where $S_{CO} \Delta v$ is the limit on the velocity integrated line flux 
in Jy~km~s$^{-1}$, $\nu_{obs}$ is the observing frequency in GHz, 
and $D_L$ is the luminosity distance in Mpc.
The choice of cosmological parameters enters in $D_L$, and 
we adopt $H_0 = 75$ km~s$^{-1}$, $\Omega = 1$ and $\Omega_{\Lambda} = 0$ 
for consistency with most work in this field. 
(An alternative cosmology with
$H_0 = 75$ km~s$^{-1}$, $\Omega = 1$ and $\Omega_{\Lambda} = 0.7$ 
results in $D_L$ larger by a factor of 1.54 for this redshift.) 
The effective linewidth is not known, but it likely falls in 
the range 150 to 550~km~s$^{-1}$ found for a large sample of 
ultraluminous galaxies in the local universe (Solomon et al. 1997).
For the $3\sigma$ flux limit obtained 
in the more sensitive part of the CO J=2--1 spectrum,
assuming a linewidth of 200~km~s$^{-1}$,
$L^{'}_{CO}(2-1) < 5.1 \times 10^{10}$ K~km~s$^{-1}$~pc$^2$.
For the $3\sigma$ flux limit obtained for CO J=5--4,
again assuming a linewidth of 200~km~s$^{-1}$, 
$L^{'}_{CO}(5-4) < 3.0 \times 10^{10}$ K~km~s$^{-1}$~pc$^2$.
If the assumed linewidth were two times larger, then 
these luminosity limits would be $\sqrt{2}$ times higher. 

Conversion of these CO luminosity limits to molecular gas mass 
limits is fraught with uncertainties.  But a simple conversion factor 
from CO luminosity to H$_2$ mass is commonly taken to be 
$4.5~M_{\odot}$ (K~km~s$^{-1}$~pc$^2$)$^{-1}$,
the value determined for Milky Way molecular clouds 
(Sanders, Scoville \& Soifer 1991). There is evidence from 
comparisons of luminosity based mass estimates with dynamical mass estimates 
that the conversion factor may be perhaps five times lower 
in ultraluminous objects (Downes \& Solomon 1998). 
Additional corrections of order unity are also needed to account properly 
for excitation from the elevated cosmic background radiation at high redshift. 
Adopting the Galactic conversion factor for CO J=2--1 line luminosity 
gives a limit on the {\em cold or diffuse} molecular gas mass 
of $\sim2.3\times10^{11}~M_{\odot}$ in the SDSS~1044-0125 system. 
Using the same conversion factor for the CO J=5--4 line luminosity
gives a limit on the {\em warm and dense} molecular gas mass of
of $\sim1.3\times10^{11}~M_{\odot}$ in the SDSS~1044-0125 system. 

These mass limits are comparable to the mass indicated from 
the detection of CO J=5--4 emission from some $z>4$ quasars, 
including at least two thought 
not to be amplified by gravitational lensing.
In particular, observations of CO J=5--4 emission from
BR1202-0725 at $z=4.7$ (Omont al. 1996, Ohta et al. 1996) 
and BRI1335-0417 at $z=4.4$ (Guilloteau et al. 1997)
indicate molecular gas masses in excess of $10^{11}$~M$_{\odot}$ 
(adjusted for the cosmology and CO to H$_2$ conversion factor
adopted here).
There is no clear physical argument to explain why some quasar 
environments show CO emission at this sensitivity level while 
others do not (Guilloteau et al. 1999).
In any case, the CO J=2--1 and J=5--4 luminosity limits suggest that 
the environment of SDSS~1044-0125 does not possess an enormous 
mass reservoir of either low excitation or high excitation molecular gas. 

The CO J=2--1 limit is comparable to the amount of molecular gas 
detected toward the lensed quasar APM~08279+5255, where 
Papadopoulos et al. (2001) found several CO J=2--1 emission features with 
total luminosity $6.6\pm3.1 \times 10^{11}$ K~km~s$^{-1}$~pc$^2$ 
attributed to (unlensed) molecular gas rich companion galaxies to the
quasar host. For the SDSS~1044-0125 observations, such features 
would have been contained within one synthesized beam (together with 
any nuclear emission). The luminosity limit suggests that no comparable 
population of nearby massive companions is present. 
Moreover, no significant CO J=2--1 emission features are found 
within the entire field of view that spans $\sim1.4$~Mpc, which 
suggests that such massive cold molecular gas concentrations are rare.
Observations of SDSS~1044-0125 with
better sensitivity are needed to explore whether smaller but still 
significant concentrations of low excitation molecular gas are 
present in the environment of this high redshift quasar.

\acknowledgments
The work was supported in part by NSF grant AST-21795 to the University
of California.  We thank Leo Blitz for scheduling additional 
observations on this project after revision of the quasar redshift,
and Alberto Bolatto for a generous donation of observing time.

\clearpage

\clearpage

\end{document}